\documentclass{aa}  

\usepackage{graphicx}
\usepackage{subfigure} % subfiguras
\usepackage{xcolor}
\usepackage{natbib}        % required for bibliography
\usepackage{txfonts}
\usepackage{eufrak}
\usepackage{multirow}

\usepackage{calligra}
\usepackage[utf8]{inputenc}
\usepackage{amsmath}
\usepackage{amsfonts}
\usepackage{amssymb}
\usepackage{calligra}
\usepackage{calrsfs}
\usepackage[mathscr]{euscript}

\DeclareSymbolFont{cmsymbols}{OMS}{cmsy}{m}{n}
\SetSymbolFont{cmsymbols}{bold}{OMS}{cmsy}{b}{n}
\DeclareSymbolFontAlphabet{\mathcal}{cmsymbols}
\newcommand{\m}[1]{\mathcal{#1}}
\providecommand{\dparcial}[2]{\frac{\partial #1}{\partial #2}}
\usepackage{hyperref}
\begin{document} 
   \title{A method to deconvolve stellar rotational velocities III}
   \subtitle{The probability distribution function via Maximum Likelihood utilizing Finite Distribution Mixtures}
   \author{
    R. Orellana \inst{1,2}
   	\and
    P. Esc\'arate \inst{3,4,5}
    \and
    M. Cur\'e \inst{6}
    \and
    A. Christen \inst{7}
    \and
    R. Carvajal \inst{1}
    \and
    J.C. Ag\"uero \inst{1}}
   \institute{
    Electronics Engineering Department, Universidad T\'ecnica Federico Santa Mar\'ia, Valpara\'iso, Chile
    \and
    Universidad de Los Andes, M\'erida, Venezuela.
    \and
    Instituto de Electricidad y Electr\'onica, Facultad de Ciencias de la Ingenier\'ia, Universidad Austral, Valdivia, Chile.\\
    \email{pedro.escarate@uach.cl}
    \and
    Large Binocular Telescope Observatory, Steward Observatory, Tucson, AZ 85546, USA
    \and
    N\'ucleo Milenio de Formaci\'on Planetaria - NPF, Chile
    \and
    Instituto de F\'isica y Astronom\'ia, Universidad de Valpara\'iso, Chile.
    \and
    Instituto de Estad\'istica, Pontificia Universidad Cat\'olica de Valpara\'iso, Chile.    
    }
 
  \abstract
   {}
   {The study of accurate methods to estimate the distribution of stellar rotational velocities is important for understanding many aspects of stellar evolution. From such observations we obtain the projected rotational speed $(v\sin i)$ in order to recover the true distribution of the rotational velocity. To that end, we need to solve a difficult inverse problem that can be posed as a Fredholm integral of the first kind}
   {In this work we have used a novel approach based on Maximum likelihood (ML) estimation to obtain an approximation of the true rotational velocity probability density function expressed as a sum of known distribution families. In our proposal, the measurements have been treated as random variables drawn from the projected rotational velocity probability density function. We analyzed the case of Maxwellian sum approximation, where we estimated the parameters that define the sum of distributions.}
   %We apply Monte-Carlo simulations to analyze the performance of the algorithm proposed.
   {The performance of the proposed method is analyzed using Monte Carlo simulations considering two theoretical cases for the probability density function of the true rotational stellar velocities:  i) an unimodal Maxwellian probability density distribution and ii) a bimodal Maxwellian probability density distribution. The results show that the proposed method yielded more accurate estimates in comparison with the Tikhonov regularization method, especially for small sample length $(N=50)$. Our proposal was evaluated using real data from three sets of measurements, and our findings were validated using three statistical tests.}
   {The ML approach with Maxwellian sum approximation is a accurate method to deconvolve the rotational velocity probability density function, even when the sample length is small ($N= 50$).}

\maketitle
%
%-------------------------------------------------------------------

\section{Introduction}
	
The estimation of the probability distribution of rotational velocities of stars is essential to describe and model many aspects of stellar evolution. From observations, it is only possible to obtain the projected velocity, $(v\sin i)$, where $i$ is the inclination angle with respect the line of sight and $v$ is the true (non-projected) rotational velocity. These measurements are assumed to be realizations of a random variable drawn from a probability density function, $f_{Y}(y|\beta)$, that satisfies the following (see e.g., \cite{Cure2014}):
\begin{equation}
f_{Y}(y|\beta) = \int p(y|x)f_{X}(x|\beta)dx.
\label{eq:fredholm_integral}
\end{equation}
where the unknown function, $f_{X}(x|\beta)$, appears under the integral sign. In our problem of interest, $f_{Y}(y|\beta)$ is the probability density function (PDF) of the available measurements, $f_{X}(x|\beta)$ is the unknown distribution, $p(y|x)$ is the conditional distribution of the projected angles and $\beta$ is a parameter vector that defines the marginal distributions. 

Typically, $f_{X}(x|\beta)$ is solved from the integral equation Eq. \eqref{eq:fredholm_integral} utilizing data $y=(y_1, \cdots, y_N)$ to estimate $f_{Y}(y|\beta)$. This corresponds to a standard solution of the Fredholm equation Eq. \eqref{eq:fredholm_integral} (see \cite{Chandrasekhar1950, Lucy1974}). There are several methods that deal with the Fredholm integral, see for example\cite{Yalcinbas, Shirin2013, Alipanah2011}. These methods solve the integral equation by approximating the unknown function in the integral via basis functions or polynomials. However, the unknown function in our problem in Eq. \eqref{eq:fredholm_integral} is a PDF. Hence, the solution of the Fredholm equation must satisfy
\begin{equation} f_{X}(x|\beta)>0, \, \forall x\quad \wedge \quad \int_{-\infty}^\infty f_{X}(x|\beta)dx = 1,
\label{eq:constraints}
\end{equation}
which is not always the case for basis functions and polynomials. Therefore, it is necessary to develop mathematical methods to actually deconvolve the measured projected velocity in order to determine the true PDF of the rotational velocity. In addition, it is necessary to directly estimate the PDF for easy handling and for the analysis of important properties of the distribution (e.g., mean, mode, kurtosis, etc.).

A standard assumption used for $p(y|x)$ in the deconvolution problem in  Eq. \eqref{eq:fredholm_integral} is that the stellar rotational axes are uniformly distributed over the unit sphere. Using this assumption, \cite{Chandrasekhar1950} studied the Fredholm integral (Eq. \eqref{eq:fredholm_integral}) that describes the probability distribution of true $(v)$ and apparent $(v\sin i)$ rotational velocities, obtaining a formal solution to it that is proportional to the derivative of an Abel's integral. Nevertheless, this method is not usually applied, because the differentiation of the formal solution can yield misleading results due to numerical problems related to the derivative of the Abel's integral. To circumvent this problem, in \cite{Cure2014} the cumulative distribution function (CDF) from a set of samples of projected rotational velocities, $(v\sin i)$, was obtained using a novel single step method. Although the CDF provides information of the distribution of the speed of rotation, it is necessary to obtain the PDF for easy handling and for the direct estimation of certain properties of the distribution (e.g., the maximum, its symmetry, etc).
	
On the other hand, in \cite{Christen2016} a method to estimate the PDF from the Fredholm integral (Eq. \eqref{eq:fredholm_integral}) by means of the Tikhonov regularization method (TRM) was obtained.  Even though regularization methods are techniques widely used to deconvolve inverse problems such as image processing, geophysics and machine learning (see \cite{Bouhamidi2007}, \cite{Fomel2007}, \cite{Deng2013}), they do not guarantee that the solution is a PDF.%, as explained before.

In this paper we propose a novel method to obtain the maximum likelihood (ML) estimate of the parameters that define the PDF of the rotational velocity written as a Maxwellian Sum Approximation (MSA). The main idea is based on \cite{Carvajal2018}, where a general estimation algorithm is developed using data augmentation. In this work we specialize the proposal in \cite{Carvajal2018} for a MSA to estimate $f_X(x|\beta)$ from Eq. \eqref{eq:fredholm_integral}.

The main departure from \cite{Christen2016} and the standard solutions for the integral equation is that in our approach, the measurements are treated as realizations of the PDF that defines the projected rotational velocity instead of evaluations of the PDF. Thus, the observed samples are used as realizations of a known parametric random variable ($y$), belonging to a known parametric family that is a mixture (see Eq. \eqref{eq:fredholm_integral}). We use the ML method (estimators with good statistical properties) to estimate the parameters, which are calculated without directly estimating $f_Y(y)$. Other methods need to perform an approximation in order to use a Kernel density estimator (KDE) to estimate the projected rotational speed density (see e.g., \cite{Cure2014} or \cite{Christen2016}). This implies that, since we obtain the ML estimates of the parameters that define the unknown PDF, our solution is, in general, well-conditioned (more data points than unknown parameters). This is due to the fact that, in general, the length of samples will be greater than the number of unknown parameters (e.g., four parameters for a Maxwellian sum approximation with two terms).

This article is structured as follows: in Section 2 the problem of interest is described. In Section 3 we provide the mathematical description of the method for the attainment of the ML estimate of the stellar rotational velocities using MSA. In Section 4, Monte Carlo simulations are presented to show the benefits of this method, comparing the results with TRM algorithm proposed by \cite{Christen2016}. In Section 5, real samples of a stellar cluster are deconvolved using the ML estimation algorithm proposed in this work. A comparison between the estimations from our method and TRM is presented. In the final section, our conclusions and future work are shown. Finally, the proofs of our results and the details of the expressions that explain the development of this work are shown in the Appendix.

%--------------------------------------------------------------------
\section{Maximum likelihood approach}

Many inverse problems in physics and astronomy are given in terms of the Fredholm integral (Eq. \eqref{eq:fredholm_integral}) of the first kind (\cite{Lucy1974}, \cite{Hansen2010}).  In \cite{Cure2014} a method to deconvolve the inverse problem given by the Fredholm integral, Eq. \eqref{eq:fredholm_integral} is developed, obtaining the CDF for stellar rotational velocities extending the work of \cite{Chandrasekhar1950}. Assuming an uniform distribution of stellar axes over the unit sphere, this integral equation reads as follows (see \cite{Cure2014} for more details):
\begin{equation}
f_{Y}(y|\beta) = \int_{y}^{\infty}\underbrace{\frac{y}{x\sqrt{x^2-y^2}}}_{p(y|x)}f_{X}(x|\beta)dx,
\label{eq:integral_y_infnty}
\end{equation}
where $x = v$, is the true rotational speed, $y=x \sin i$, is the projected rotational speed, and $i$, is the (unknown) inclination angle. Furthermore, $f_{Y}(y|\beta)$ represents the PDF of projected rotational velocities and $f_{X}(x|\beta)$ is the density of true rotational velocities. The function $p(y|x)$ in this integral is related to the distribution of projected angles (\cite{Cure2014}). We note that the conditional distribution $p(y|x)$ in Eq.\eqref{eq:integral_y_infnty} is valid for an isotropic stellar rotational axes distribution.

The distribution of rotational velocities has been studied in the literature, providing strong evidence for the occurrence of Gaussian and Maxwellian distributions in astrophysical systems. \cite{Deutsch1970,Gaige1993} proved (analytically) that a Maxwellian distribution corresponds to the rotational speed distribution when the distribution of stellar axes is uniform distributed over the unit sphere (random axes orientations). The work in \cite{Chandrasekhar1950} revived the suggestion of van Diem that proposed a double Gaussian distribution to describe the rotational speed of stars, presenting an analysis for B to G0 stars in the Pleiades cluster. For non-Gaussian statistics, \cite{Carvalho2009} prove that the Tsallis distribution (with the $k$ parameter) corresponds to the distribution of rotational velocities, and in the limit case when $k \to 0$, the Maxwellian distribution is recovered. In addition, it is known that for a similar problem, the mass distribution of exoplanets is described by a mixture of two Gaussian distributions in the logarithm of the planet mass (see e.g., \cite{Malhotra2015}). However,  the distribution $f_X(x|\beta)$ is usually unknown (not just the parameters, but also the parametric family). In such cases, an adequate alternative is to approximate the PDF using a distribution sum approximation (DSA) with a known and suitable PDF as a basis. This concept arises from the Wiener approximation theorem \cite{ref:Achieser}, which, in simple terms, states that any function $\mathfrak{f}(\cdot)$ can be approximated by a linear combination of translations of another given function,  $\mathfrak{g}(\cdot)$, if  the Fourier transform of $\mathfrak{g}(\cdot)$ is not equal to zero in the domain of interest. A common choice for a basis to form a DSA is the Gaussian distribution \cite{Alspach1972}, which satisfies the Wiener approximation theorem. In our problem, we consider a Maxwellian sum approximation, since it has been suggested that the distribution of stellar rotational velocities can be modelled by a Maxwellian distribution \cite{Chandrasekhar1950},  \cite{Deutsch1970,Gaige1993}. 
%\begin{remark}
%An MSA does not approximate every PDF, unless a translation term is included in the Maxwellian distribution. Nevertheless, as shown in the following sections, \textbf{a} MSA, even without translation, may provide an adequate fit to the PDF of the stellar rotational velocities. \eor
%\end{remark}
We note that an MSA does not approximate every PDF, unless a translation term is included in the Maxwellian distribution. Nevertheless, as shown in the following sections, a MSA, even without translation, may provide an adequate fit to the PDF of the stellar rotational velocities.

The unknown true rotational velocities distribution function can be expressed as the following DSA:
\begin{equation}
\label{eq:fx_DSA}
f_{X}(x|\beta)\approx \sum_{j=1}^{K}\lambda_j g(x|\theta_j),
\end{equation} 
%subject to
%\begin{equation}
%\sum_{j=1}^{K}\lambda_j=1
%\end{equation}
where $g(x|\theta_j)$ represents a PDF characterized by the parameter $\theta_j$, where $\lambda_j$ is the weight of the corresponding $jth$ distribution and $K$ represents the number of distributions used to approximate the PDF subject to $\sum_{j=1}^{K}\lambda_j=1$. Thus, the parameter vector to estimate is given by
\begin{equation}
\beta = [\underbrace{\lambda_1, \theta_1}_{\beta_1},\dots, \underbrace{\lambda_K, \theta_K}_{\beta_K}].
\end{equation}
If we assume that the available data $y_t$ ($t=1,\dots,N$) are independent and identically distributed random variables, we obtain for $y=(y_1, \cdots, y_N)$: %and $x_T$ 
\begin{equation}
f_{Y}(y|\beta) = \prod_{t=1}^N f_{Y}(y_t|\beta),
\end{equation}
where
\begin{align}
\label{int_pyt_GMM}
f_{Y}(y_t|\beta)&=\sum_{j=1}^{K}\lambda_j f_j(y_t|\theta_j),\\
\label{eq_fyt_theta}
f_j(y_t|\theta_j)&=\int_{y_t}^{\infty} p(y_t|x_t)g(x_t|\theta_j)dx_t.
\end{align}
Then, the likelihood function can be expressed as
\begin{equation}
\m L_N(\beta)=f_{Y}(y_{1:N}|\beta) =\prod_{t=1}^{N}\sum_{j=1}^{K}\lambda_jf_j(y_t|\theta_j).
\end{equation}
The log-likelihood function $\ell_N(\beta)=log (L_N(\beta))$ reads
\begin{equation}
\ell_N(\beta)=\sum_{t=1}^{N}\log\left[\sum_{j=1}^{K}\lambda_j f_j(y_t|\theta_j)\right].
\label{eq:log_likelihood}
\end{equation}
We note that the expression in Eq. \eqref{eq_fyt_theta} can be understood as a generalized probability density function (see, e.g., \cite{Degroot2004}). We develop our estimation algorithm in the following section based on this interpretation.

%=============================================
% Maximum Likelihood estimation using Maxwellian sum approximation
%=============================================
\section{Maximum likelihood estimation using Maxwellian sum approximation}
In this section we develop the ML estimation algorithm when modelling $f_{X}(x|\beta)$ as a Maxwellian sum approximation (MSA). This work is an extension of the estimation procedure proposed in  \cite{Carvajal2018} and in \cite{Orellana2018}. In particular, in this work we solve the problem of interest (Eq. \ref{eq:integral_y_infnty}) considering a discrete Maxwellian mixture.

In \cite{Carvajal2018} an estimation algorithm was shown for a general class of problems with data augmentation, based on the Expectation-Maximization (EM) algorithm \citep{Dempster1977}. They consider a general optimization problem that can be tailored to solve the problem in Eq. \eqref{eq:integral_y_infnty} to estimate the parameters that define the approximation of the true rotational velocities (e.g: MSA). Inspired by the EM algorithm, an optimization problem using an auxiliary function is defined, and then an iterative algorithm is obtained.

It is worth noting that the estimation method we are proposing is a parametric one, based on the ML estimation principle, giving a set of parameters that define the PDF of true rotational velocities using a sum of $K$ Maxwellian distributions, to describe a sample of data. This is an important difference with respect to the TRM proposed by \cite{Christen2016} where they present a non-parametric estimation approach; from the projected rotational velocities applying a KDE used by \cite{Silverman1986} discretizing the Fredholm integral Eq. (\ref{eq:integral_y_infnty}) and obtained the Tikhonov regularization solution.

For the special case of the distribution of a sample of stellar rotational speeds,\cite{Deutsch1970,Gaige1993} demonstrated that the density $f_{X}(x|\beta)$ is given by a Maxwellian distribution with dispersion $\sigma$. In Appendix \ref{appendix_B} we show the details of our MSA formulation. The optimization of the auxiliary function with respect to vector of parameters $\beta$ is shown as follows.

Considering $g(x|\theta_j)$ as a Maxwellian PDF (Eq. \eqref{eq:fx_DSA}). The Maxwellian PDF is defined by
\begin{equation}
\label{eq:Maxwellian_dist}
\phi_{M}(x;\sigma_j)=\sqrt{\frac{2}{\pi}}\frac{x^2}{\sigma_j^3}\exp\left\{-\frac{x^2}{2\sigma_j^2}\right\},
\end{equation}
where $x>0$ and $\sigma_j$ is the dispersion parameter that defines de PDF. Then, we can define the following function:
\begin{equation}
K_{M}(x_t,\beta_j)=\lambda_j\phi_{M}(x_t;\sigma_j),
\end{equation}
where $\beta_j=\left\{\lambda_j,\sigma_j\right\}$. The corresponding optimization problem is solved iteratively. Thus, assuming that the estimates of the parameters at the $m$th iteration are available, the vector of parameters $\hat{\beta}^{(m+1)}$ that optimize the auxiliary function (Eq. \ref{eq_Estep}) is now given by
\begin{align}
\label{eq:lamda_m1_MSA}
\hat{\lambda}_j^{(m+1)}&=\frac{\mathcal{P}_M(x_t,\hat{\beta}_j^{(m)})}{\sum_{l=1}^{K}\mathcal{P}_M(x_t,\hat{\beta}_l^{(m)})},\\
%\end{align}
%\begin{equation}
\label{eq:sigma_m1_MSA}
\hspace{5mm} \hat{\sigma}_j{}^{(m+1)}&=\left[\frac{\mathcal{S}_M(x_t,\hat{\beta}_j^{(m)})}{3\mathcal{P}_M(x_t,\hat{\beta}_j^{(m)})}\right]^{\frac12},
\end{align}
where
\begin{align}
\label{Pxt_beta_MSA}
\mathcal{P}_M(x_t,\hat{\beta}_j^{(m)})&=\sum_{t=1}^{N}\int_{y_t}^{\infty}\frac{K_M\left(x_t,\hat{\beta}_j^{(m)}\right)}{\mathcal{V}_{M_t}\left(\hat{\beta}^{(m)}\right)}d\mu(x_t),\\
%\end{equation}
%\begin{equation}
\label{Sxt_beta_MSA}
\mathcal{S}_M(x_t,\hat{\beta}_j^{(m)})&=\sum_{t=1}^{N}\int_{y_t}^{\infty}x_t^2\frac{K_M\left(x_t,\hat{\beta}_j^{(m)}\right)}{\mathcal{V}_{M_t}\left(\hat{\beta}^{(m)}\right)}d\mu(x_t),\\
%\end{equation}
%and
%\begin{equation}
\mathcal{V}_{M_t}(\hat{\beta}^{(m)})&=\sum_{j=1}^{K}\int_{y_t}^{\infty}K_{M}(x_t,\hat{\beta}_j^{(m)})d\mu(x_t).
\label{Vxt_beta_MSA}
\end{align}
In Appendix \ref{appendix_B} we demonstrate the optimization procedure to obtain the MSA. We summarise the proposed iterative algorithm with MSA as follows:
\begin{enumerate}[i)]
	\item Set a number of $K$ distributions for MSA.
	\item Obtain an initial guess $\hat{\beta}_j^{(0)}$ for $j=1,\dots,K$.
	\item Set $m=0$.
	\item Compute the integrals in Eq. \eqref{Vxt_beta_MSA}, \eqref{Pxt_beta_MSA} and \eqref{Sxt_beta_MSA}.
	\item Compute $\hat{\beta}^{(m+1)}$ from Eq. \eqref{eq:lamda_m1_MSA} and \eqref{eq:sigma_m1_MSA}.  
	\item Set $m=m+1$ and go back to step iv) until convergence is achieved.
\end{enumerate}
	
In iterative optimization algorithms, it is customary to define one or more stopping criteria, such as a maximum number of interactions and/or when the error reaches a certain value. In this work the stopping criterion is defined by the normalized relative error ${\Vert\hat{\beta}^{(m)}-\hat{\beta}^{(m-1)}\Vert}/{\Vert\hat{\beta}^{(m)}\Vert}$ reaching the arbitrary tolerance of $10^{-6}$. That is, 
\begin{equation}
\label{rel-err}
\hspace{5mm}  \frac{\Vert\hat{\beta}^{(m)}-\hat{\beta}^{(m-1)}\Vert}{\Vert\hat{\beta}^{(m)}\Vert} < 10^{-6}
\end{equation}

%\begin{remark}
%It is possible to utilize a Gaussian sum approximation instead of \textbf{a} MSA. In fact, it was shown in \cite{Lo1972} that any PDF can be approximated as closely as desired by a density of the form of Gaussian sum approximation for some finite components of the mixture. However, in our experience, for the problem of interest (true rotational velocities), we need a large number of Gaussian components to obtain an adequate agreement with the true distributions, avoiding the truncation of the estimated  PDF for lower velocities. Hence, we focus on utilizing \textbf{a} MSA, comparing the performance of our proposal against \textbf{TRM method proposed by \cite{Christen2016})}. \eor
%\end{remark}
We note that it is possible to utilize a Gaussian sum approximation instead of a MSA. In fact, it was shown in \cite{Lo1972} that any PDF can be approximated as closely as desired by a density of the form of Gaussian sum approximation for some finite components of the mixture. However, in our experience, for the problem of interest (true rotational velocities), we need a large number of Gaussian components to obtain an adequate agreement with the true distributions, avoiding the truncation of the estimated  PDF for lower velocities. Hence, we focus on utilizing a MSA, comparing the performance of our proposal against TRM method proposed by \cite{Christen2016}). 

%===========================================0
% simulations
%============================================
%\vspace{-11mm}
\section{Monte Carlo simulations}
In this section we present the results of Monte Carlo (MC) numerical simulations to assess the performance of ML estimation using MSA. %GSA and MSA%.
 A comparison is made using the TRM proposed by \cite{Christen2016}. We simulated the following cases:
\begin{enumerate}[a)]
	\item {\it{Unimodal distribution}}: we chose the density of true rotational velocities $f_{M}(x|\sigma)^{\text{(True)}}$ as
	\begin{equation}
	\label{eq:unimodal_true}
	f_{\text{M}}(x|\sigma)^{\text{(True)}}=\sqrt{\frac{2}{\pi}}\frac{x^2}{\sigma^3}\exp\left\{-\frac{x^2}{2\sigma^2}\right\},\;\;\;\;\;\;\;\;x>0
	\end{equation}
	where dispersion parameter $\sigma=8$, which is the same distribution used in \cite{Cure2014} and \cite{Christen2016}. 
	
	\item {\it{Bimodal distribution}}: for a mixture of two Maxwellian distributions, the PDF reads
\begin{align}
\label{eq:Bimodal_true}
f_{\text{2M}}(x|\beta)^{\text{(True)}}&=\sqrt{\frac{2}{\pi}}x^2\left( \frac{\lambda_1}{\sigma_1^3}\exp\left\{-\frac{x^2}{2\sigma_1^2}\right\}\right.+\\ \nonumber
& \left.\frac{\lambda_2}{\sigma_2^3}\exp\left\{-\frac{x^2}{2\sigma_2^2}\right\}\right) ,\;\;\;\;\;\;\;\;\;\;\;\;\;\;\;\;x>0
\end{align}
where $\beta = [\lambda_1,\sigma_1,\lambda_2,\sigma_2]$, the dispersion parameters are: $\sigma_1=5$ and $\sigma_2=15$, and the mixing weights are $\lambda_1=0.7$ and $\lambda_2=0.3$, which corresponds to the same distribution used in \cite{Christen2016}.
\end{enumerate}
In \cite{Cure2014} it is shown an explicit analytic solution from the Fredholm integral Eq. (\ref{eq:integral_y_infnty}) for these cases, obtaining
\begin{align}
\label{fy_Mtrue}
f_{\text{M}}(y|\sigma)^{\text{(True)}}&=\frac{y}{\sigma^2}\exp\left\{-\frac{y^2}{2\sigma^2}\right\},\\
%\end{align}
%\begin{equation}
\label{fy_2Mtrue}
f_{\text{2M}}(y|\beta)^{\text{(True)}}&=\lambda_1\frac{y}{\sigma_1^2}\exp\left\{-\frac{y^2}{2\sigma_1^2}\right\}+\lambda_2\frac{y}{\sigma_2^2}\exp\left\{-\frac{y^2}{2\sigma_2^2}\right\},
\end{align}
The synthetic data $y_{1},y_2,\dots,y_N$, for both cases, is generated using the Slice Sampler (see e.g.,\cite{neal2003}). The simulation setup is as follows:
\begin{enumerate}[(1)]
	\item Three sample lengths $N = 50, 500, 2000$ are considered.
	\item The MSA algorithms with three values for the number of distributions $K = 1, 2, 3$ is considered. The stopping criterion is given by Eq. \eqref{rel-err}.
    \item The TRM algorithm is set following the procedure in \cite{Christen2016}.
	\item The number of MC simulations is $n_{\text{MC}}=100$.
%	\item The stopping criterion is given by Eq. \eqref{rel-err}
\end{enumerate}
\vspace{-3mm}
\subsection{Estimation using the Tikhonov regularization method}
\begin{figure*}[tb]
	\centering
	\subfigure{\includegraphics[width=0.3\textwidth]{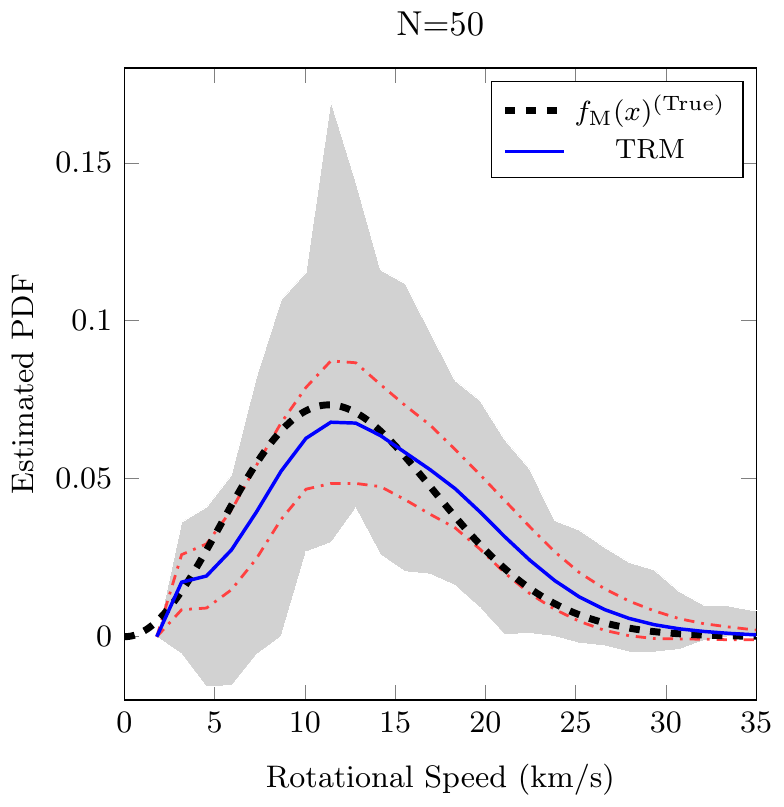}}
	\hfill
	\setcounter{subfigure}{0}
	\subfigure[]{\includegraphics[width=0.3\textwidth]{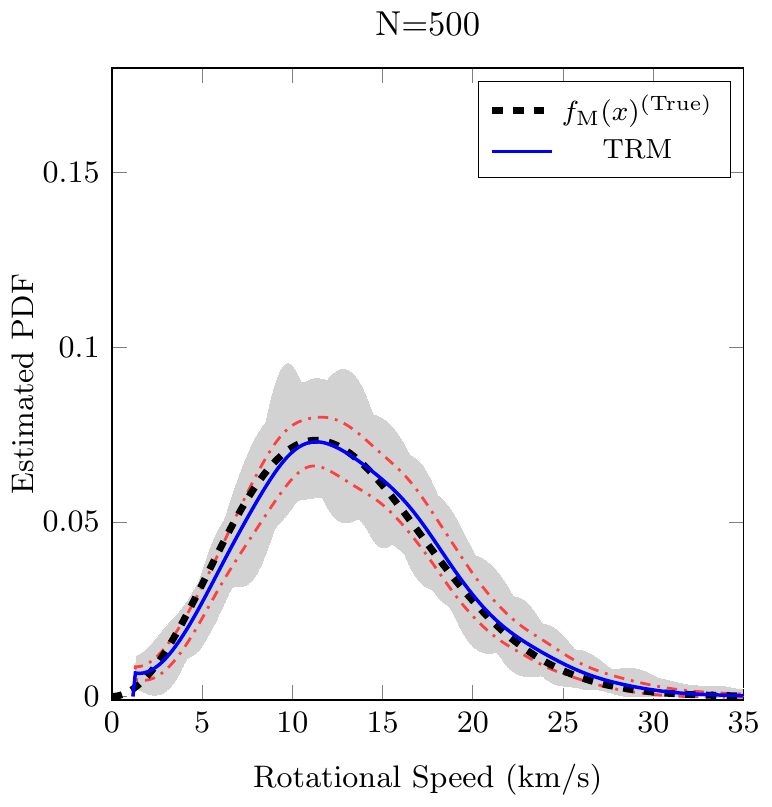}}
	\hfill
	\subfigure{\includegraphics[width=0.3\textwidth]{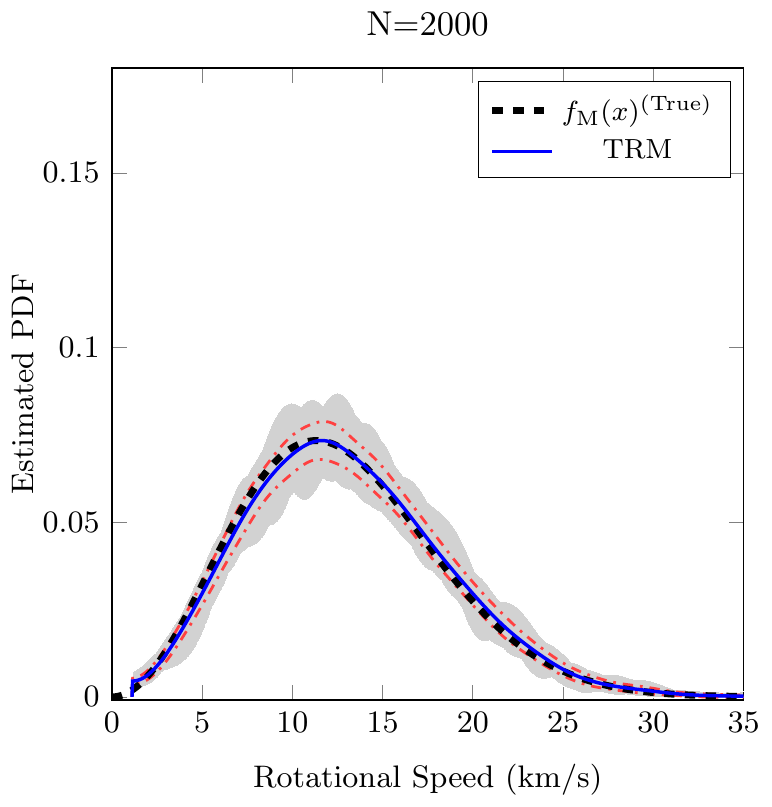}}
	\hfill
	\setcounter{subfigure}{0}
	\subfigure{\includegraphics[width=0.3\textwidth]{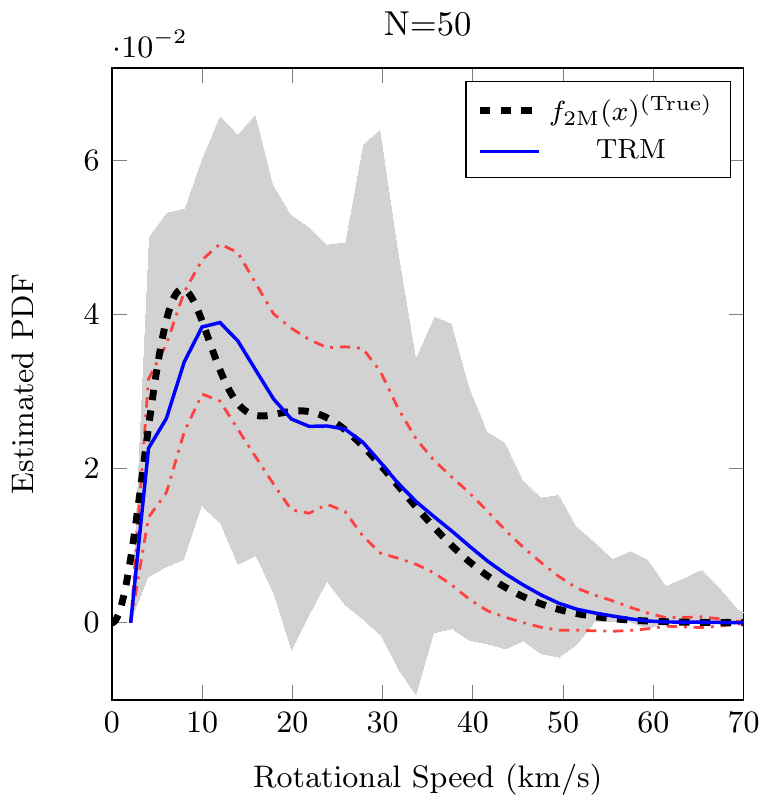}}
	\hfill
	\subfigure[]{\includegraphics[width=0.3\textwidth]{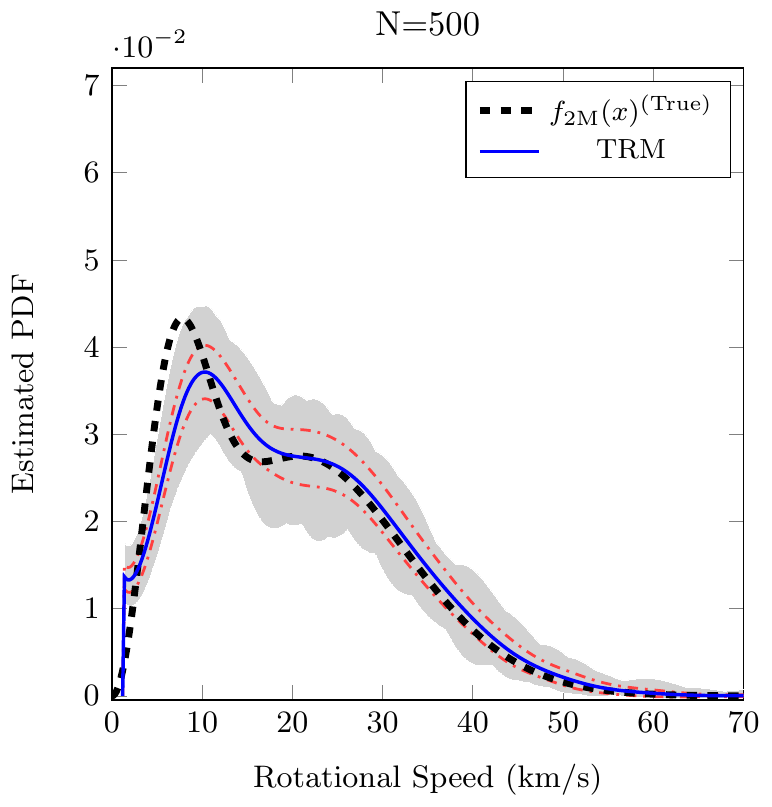}}
	\hfill
	\subfigure{\includegraphics[width=0.3\textwidth]{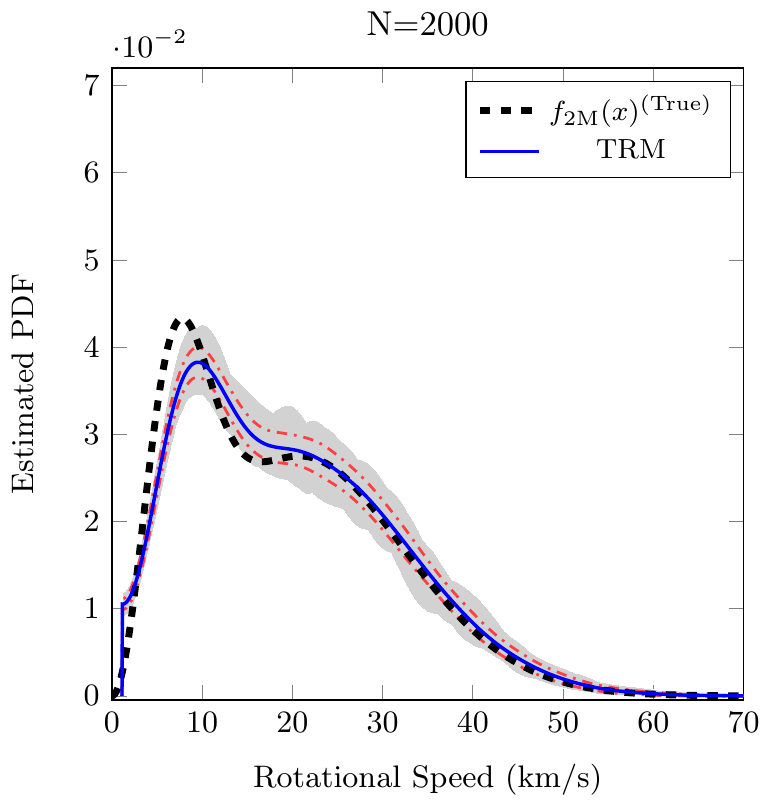}}
	%	\subfigure[]{\includegraphics[width=1\textwidth,height=1\columnwidth,keepaspectratio,trim=38mm 0.78mm 28mm 0.7mm, clip]{Fig1_a_V10.eps}}
	%	\subfigure[]{\includegraphics[width=1\textwidth,height=1\columnwidth,keepaspectratio,trim=38mm 0.78mm 28mm 0.7mm, clip]{Fig1_b_V10.eps}}
	\caption{Monte Carlo simulation results to true rotational velocities PDF estimation: (a) Using TRM algorithm for unimodal Maxwellian distribution. (b) Using TRM algorithm for bimodal Maxwellian distribution. The black line represents the true rotational velocities PDF.  The blue line represents the average of the estimated PDF using TRM. The region between the red dashed-dotted lines corresponds to one standard deviation level of the all estimated PDFs.}
	\label{fig:MC_simulations_TRM}
\end{figure*}

	In order to estimate the performance of TRM algorithm with simulated data, we follow the procedure described in \cite{Christen2016} for each independent MC simulation considering the three sample lengths cases. The TRM algorithm performance is closely related with the PDF estimation of the projected rotational velocities using KDE \cite{Silverman1986}. They use a kernel distribution as a non-parametric representation of the PDF of projected velocities, whose behavior is sensitive to the smoothing of the function and the bandwidth (BW) value. This features control the smoothness of the resulting density distribution. Figure \ref{fig:MC_simulations_TRM} (a) shows the mean estimated PDFs of all MC simulations for unimodal distribution using the TRM algorithm. Similarly, Fig. \ref*{fig:MC_simulations_TRM} (b) (bimodal distribution) exhibits the mean estimated PDFs. The gray-shaded regions (Fig. \ref*{fig:MC_simulations_TRM} (a) and Fig. \ref*{fig:MC_simulations_TRM} (b)) represent the surrounding area in which independent MC simulations lie. Also, the region between the red dashed-dotted lines corresponds to one standard deviation level of the all estimated PDFs. We note a variability in the PDF estimation for all cases, especially for small sample data lengths of order $N = 50$. In this sense, a different setting for KDE estimation (e.g. different smoothing of the function and different BW value) would be needed for each MC simulation to improve the PDF estimation, due to the fact that the bandwidth parameter controls the smoothness of the resulting density curve (see e.g., \cite{Silverman1986}).

	We note that for small sample lengths some MC simulations might not yield a PDF, since negative values could be obtained. In addition, the true rotational velocities that is more likely to occur is wrongly estimated by $2-3$ km/s for unimodal distribution and by $5$ km/s for the bimodal distribution. Despite of the above, the mean estimated PDFs show a good agreement with the true distributions (unimodal and bimodal distributions) for larger sample lengths ($N= 500$ and $N= 2000$). We also observe that the shaded regions decrease in area, implying that the estimates for the MC simulations exhibit a smaller variance.
	
	%, setting the bandwidth value around $\text{BW} = 1.5$ for the KDE estimator. 
	%setting the bandwidth KDE parameter around $\text{BW} = 3$.
	
\vspace{-3mm}
\subsection{Estimation using Maxwellian sum approximation}
In order to evaluate the performance of the method that we propose,  we considered a MSA with $K = 1$ for the unimodal and a MSA with $K = 2$ for bimodal cases. From our simulations, the weights of  the MSA are $\lambda_j \approx 0$, $j\geq 2$, for the unimodal case, whilst  for the bimodal PDF the same is true when $j\geq 3$.
\begin{figure*}[tb]
	\centering
	\subfigure{\includegraphics[width=0.3\textwidth]{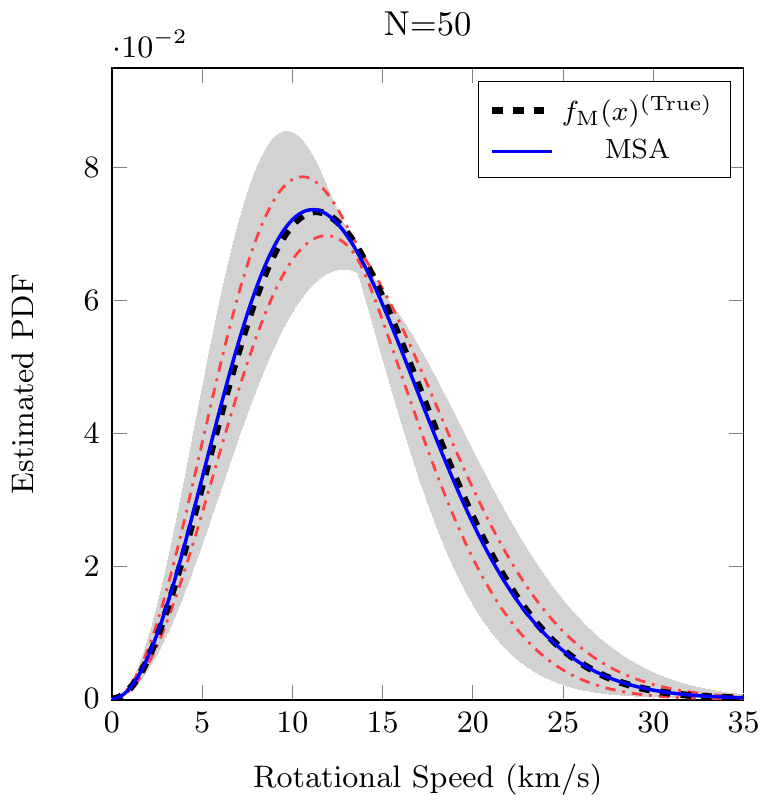}}
	\hfill
	\setcounter{subfigure}{0}
	\subfigure[]{\includegraphics[width=0.3\textwidth]{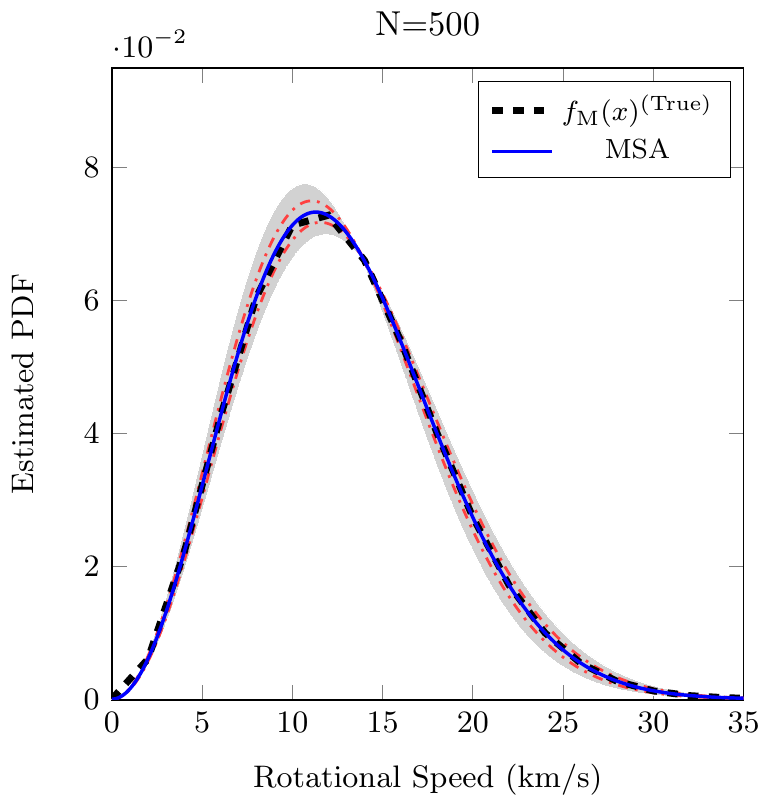}}
	\hfill
	\subfigure{\includegraphics[width=0.3\textwidth]{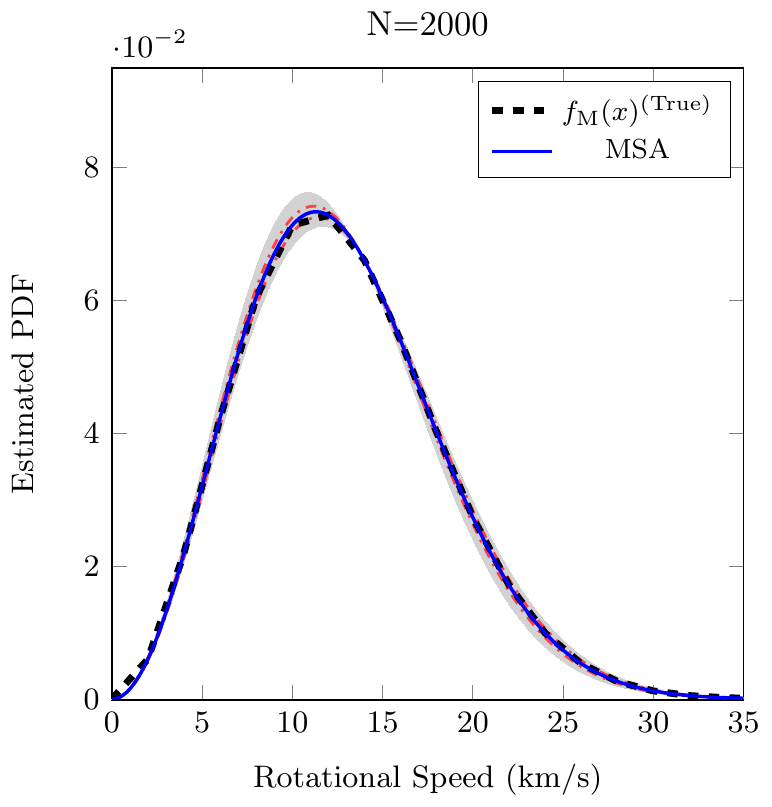}}
	\hfill
	\setcounter{subfigure}{0}
	\subfigure{\includegraphics[width=0.3\textwidth]{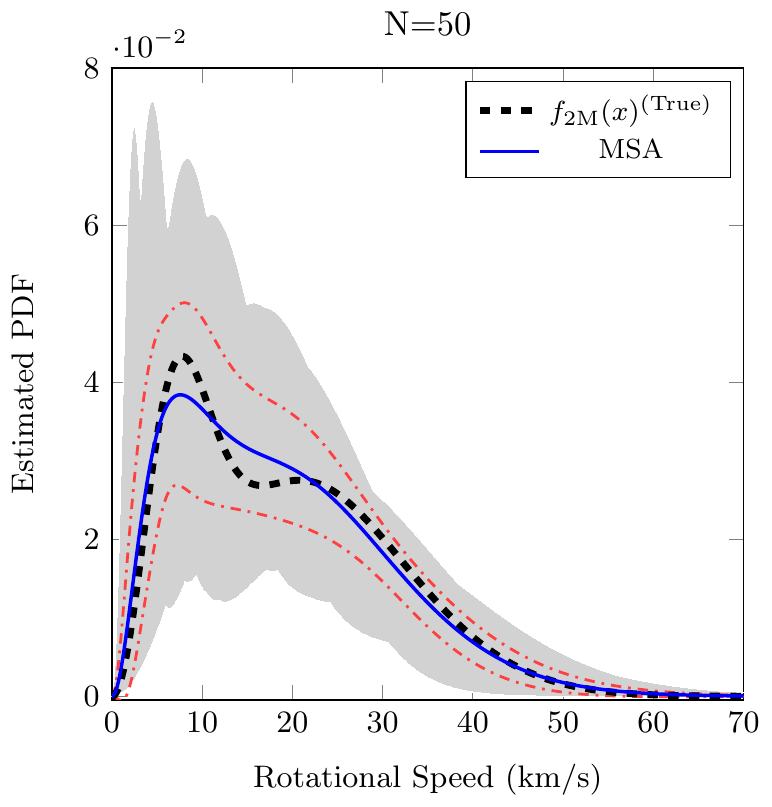}}
	\hfill
	\subfigure[]{\includegraphics[width=0.3\textwidth]{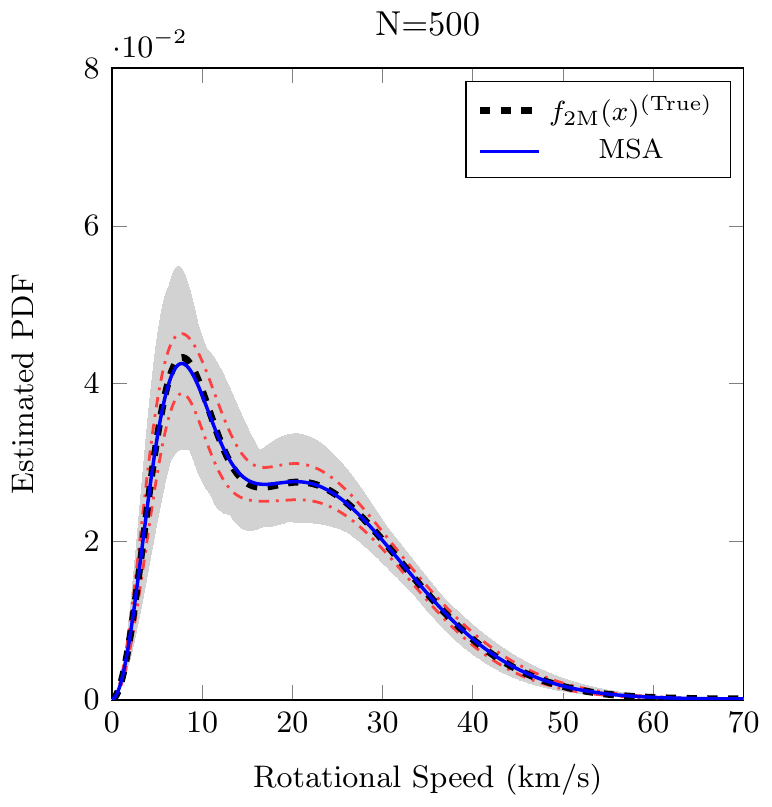}}
	\hfill
	\subfigure{\includegraphics[width=0.3\textwidth]{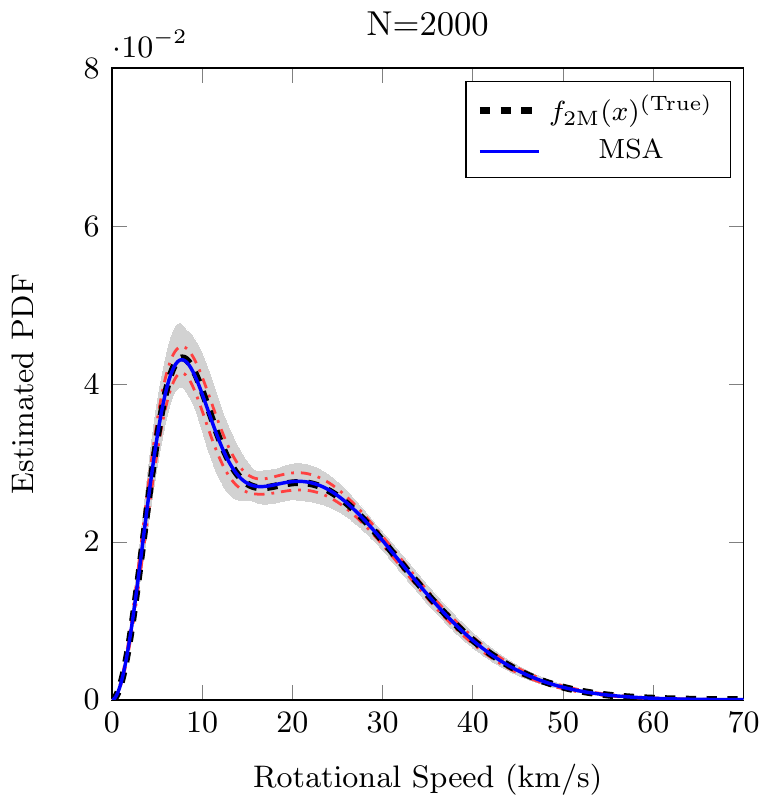}}
	%\subfigure[]{\includegraphics[width=1\textwidth,height=1\columnwidth,keepaspectratio,trim=38mm 0.78mm 28mm 0.7mm, clip]{Fig2_a_V10.eps}}
	%\subfigure[]{\includegraphics[width=1\textwidth,height=1\columnwidth,keepaspectratio,trim=38mm 0.78mm 28mm 0.7mm, clip]{Fig2_b_V10.eps}}
	\caption{Monte Carlo simulation results to true rotational velocities PDF estimation: (a) Using $K=1$ for MSA algorithm. (b) Using $K=2$ for MSA algorithm. The black line represents the true rotational velocities PDF. The blue line represents the average of the estimated PDF using MSA. The region between the red dashed-dotted lines corresponds to one standard deviation of the all estimated PDFs.}
	\label{fig:MC_simulations_DSA}
\end{figure*}
	
	Figure \ref*{fig:MC_simulations_DSA} (a) shows the true unimodal distribution (Eq. \eqref{eq:unimodal_true}) together with the mean estimated PDF of all MC simulations. The gray-shaded regions represent the surrounding area to which all independent MC simulations using the MSA algorithm lie in (Fig. \ref{fig:MC_simulations_DSA} (a) and Fig. \ref{fig:MC_simulations_DSA} (b)). Also, the region between the red dashed-dotted lines corresponds to one standard deviation level of the all estimated PDFs using our algorithm. For larger sample length ($N= 2000$), the MSA based estimation exhibits an excellent agreement between the true unimodal distribution and the mean estimated PDFs. Fig. \ref*{fig:MC_simulations_DSA} (b) shows the results for the bimodal Maxwellian distribution (Eq.\ref{eq:Bimodal_true}). On average, the agreement between the estimated PDF using MSA algorithm and the true distribution is very good, even for sample lengths of the order $N= 50$, since the mean estimated PDF is very similar to the true PDF. We note that the shaded regions show less variability of the estimated PDFs than TRM (Fig. \ref{fig:MC_simulations_TRM}(a) and Fig. \ref{fig:MC_simulations_TRM}(b)). For MSA, the variance of the estimates is higher for small sample lengths ($N= 50$) than large sample length ($N= 500,2000$) (see Fig. \ref{fig:MC_simulations_DSA} (a) and Fig. \ref{fig:MC_simulations_DSA} (b)). The initialization of the MSA estimation algorithm was set to a random initial guess. In our experience, this was good enough to provide good estimates, exhibiting good convergence properties. In Appendix \ref{appendix_B3} we show tables with the mean values and standard deviations of the estimated parameters using MSA. We observe that, in general, the estimated parameters are close to the true values for all sample lengths when the MSA algorithm is used, including sample lengths of order $N= 50$.

Following \cite{Cure2014} method to compare TRM and MSA algorithms, we calculate the mean integrated square error (MISE) to quantify the error of the estimated PDF, respect the true PDF, then we have:
\begin{equation}
\label{eq:MISE}
\text{MISE}=\frac{1}{n_{\text{MC}}}\sum_{j=1}^{n_{\text{MC}}}\left(\frac{1}{N}\sum_{i=1}^{N}\left(\hat{f}_j(x_i|\hat \beta)-f_j(x_i|\beta)\right)^2\right),
\end{equation}
where $f_j(x_i|\beta)$ represents the true PDF of rotational speeds evaluated at the $ith$ sample of the $j$th MC simulation, $\hat{f}_j(x_i|\hat \beta)$ is the estimated PDF using the MSA algorithm for the $j$th MC simulation evaluated at the $i$th sample, $n_{MC}$ is the number of MC simulations and $N$ is the sample length. In Appendix \ref{appendix_B4} we summarize the MISE values for the MC simulations for the MSA and TRM estimation. Table \ref{Tab:MISE_methods_unimodal} shows the MISE values for unimodal Maxwellian estimation ($K=1$), where the MISE for MSA algorithm is in the order of $\text{MISE}\approx 10^{-6}$. %The GSA shows a decreasing MISE when the sample length is increases. 
In Table \ref{Tab:MISE_methods_Bimodal} we show the MISE values for different sample lengths $N$ for bimodal Maxwellian distribution estimation ($K=2$). The MISE values that were obtained for MSA algorithm are all very similar, in the order of $\text{MISE}\approx 10^{-6}$. %The GSA algorithm shows that the MISE values are very similar, except for sample lengths of order $N\sim 50$.
The MC simulation show that the TRM method and the MSA method have a similar average behavior for larger samples lengths $N = 2000$ ($\text{MISE}\approx 10^{-6}$), although for small sample length  $N= 50$ the MSA method ($\text{MISE}\approx 10^{-6}$) exhibits a better estimation performance than the TRM method ($\text{MISE}\approx 10^{-5}$) in terms of average and variance . Details of the results are presented in Appendix C.
%\vspace{-4mm}

\section{Deconvolving real samples}

In this section, we perform the following: i) apply the MSA algorithm to a sample of measured $(v\sin i)$ data of stars in order to estimate the PDF of the true rotational velocities; and ii) compare the performance of this proposed algorithm with TRM proposed by \cite{Christen2016}.
\vspace{-3mm}
\subsection{Coma Berenice sample}
From the catalog \cite{Mermilliod2009}, we select the Coma Berenice cluster (Melotte 111) data which has $N=60$ values of $v\sin i > 0$ from $0\;\text{km/s}$ up to $50\;\text{km/s}$ for F-K dwarf stars.

\begin{figure*}[tb]
	\centering
	\subfigure{\includegraphics[width=0.3\textwidth]{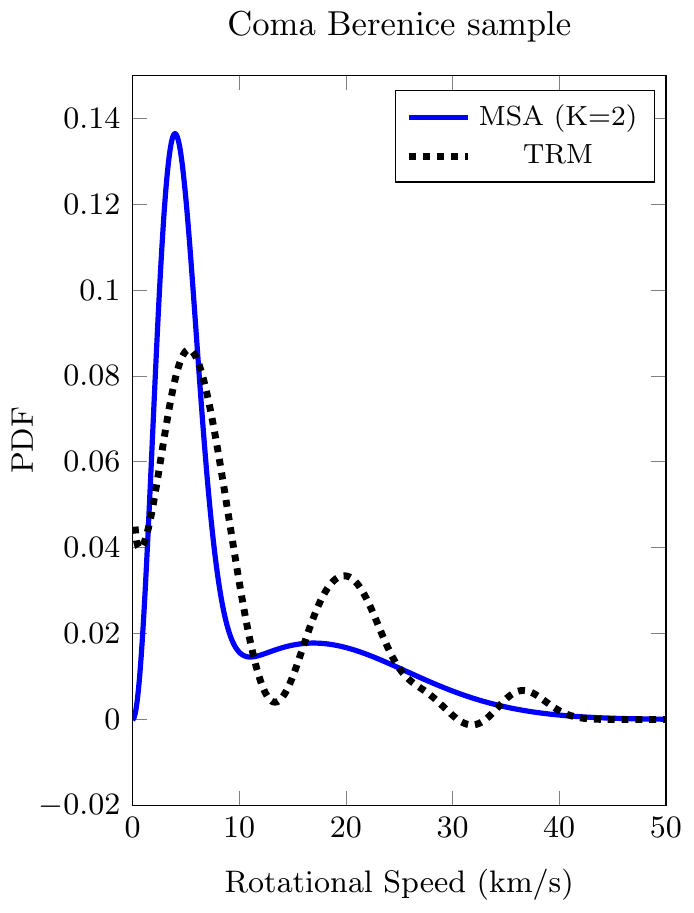}}
	\hfill
	\subfigure{\includegraphics[width=0.3\textwidth]{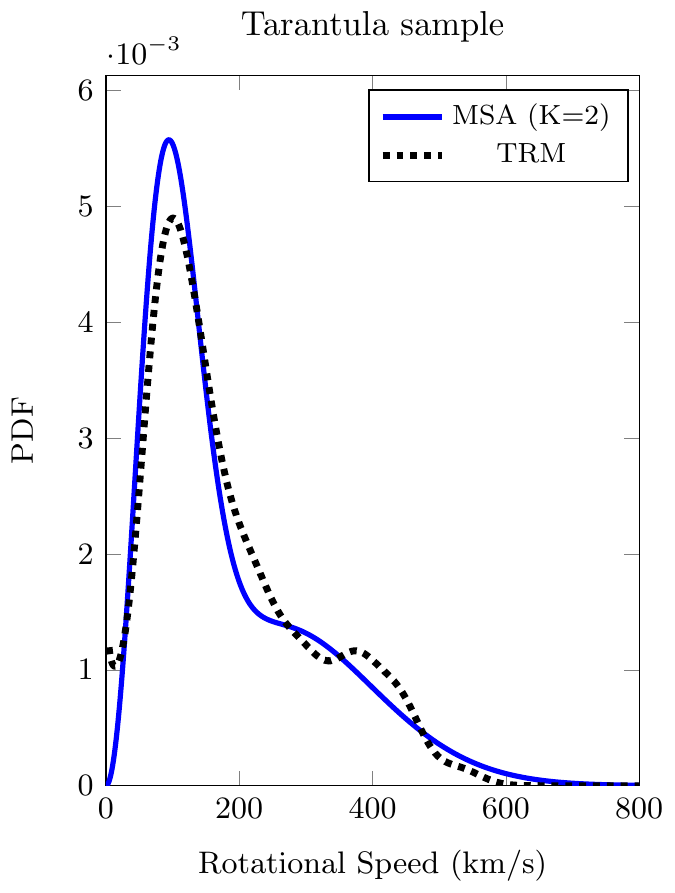}}
	\hfill
	\subfigure{\includegraphics[width=0.3\textwidth]{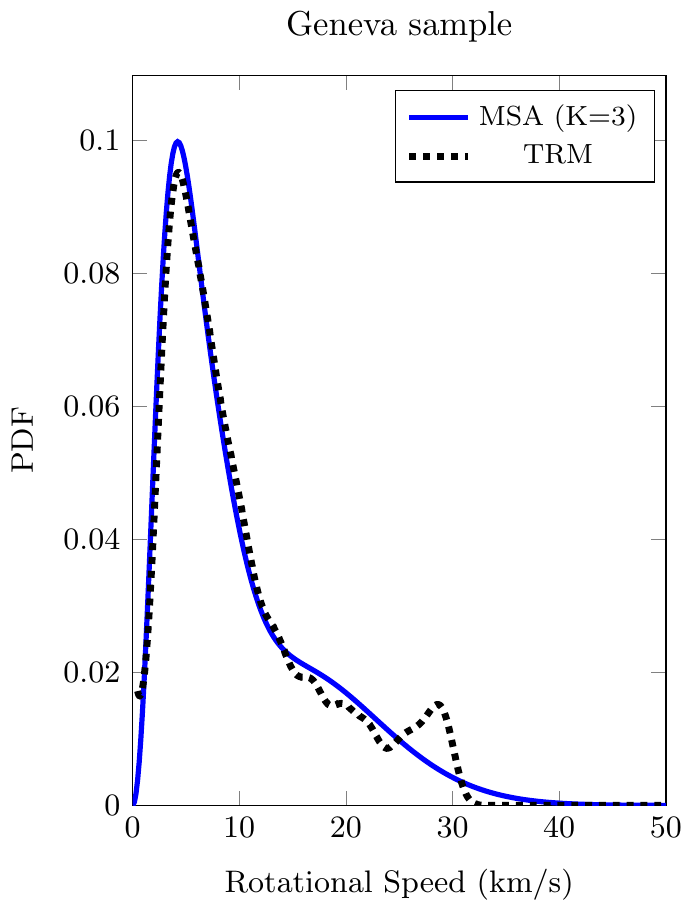}}
	%\includegraphics[width=1\textwidth,height=1\columnwidth,keepaspectratio,trim=29mm 0.78mm 29mm 1mm, clip]{Fig3_V10_2.eps}
	%\subfigure[]{\includegraphics[width=1\textwidth,height=1\columnwidth,keepaspectratio,trim=25mm 0.78mm 28mm 0.5mm, clip]{Fig3_a_V10.eps}}
	%\subfigure[]{\includegraphics[width=1\textwidth,height=1\columnwidth,keepaspectratio,trim=25mm 0.78mm 28mm 0.5mm, clip]{Fig3_b_V10.eps}}
	\caption{Estimated PDF of true rotational velocities for real samples cases using $K=2$ (Coma Berenice and Tarantula samples) and $K=3$ (Geneva sample) for MSA algorithm. The TRM algorithm is setting following \cite{Christen2016}. The blue line represents the MSA estimated PDF. The black dashed-dotted line represent the TRM estimated PDF.}
	\label{fig:Realsamples_PDFx_estimation}
\end{figure*}

The estimated parameter using MSA algorithm with one Maxwellian distribution is $\sigma_1=7.5675$. In addition, the estimated parameters using two Maxwellian distributions ($K=2$) are $\sigma_1=2.7903$, $\sigma_2=11.9939$, $\lambda_1=0.6364$ and $\lambda_2=0.3636$. Similarly, we obtain the results considering a three Maxwellian mixture distribution ($K=3$), where the estimated parameters are  $\sigma_1=2.7902$, $\sigma_2=11.9934$, $\sigma_3=11.9954$, $\lambda_1=0.6364$, $\lambda_2=0.2810$ and $\lambda_3=0.0827$. We can observe that the mixing weight of the third component ($\lambda_3$) of Maxwellian mixture distribution is close to zero. In this sense, for simplicity and clarity on the presentation,  analyzed the case for which $K=2$. In Fig. \ref{fig:Realsamples_PDFx_estimation} (left panel) we show the rotational velocity distribution using MSA algorithm with two Maxwellian mixture distribution ($K=2$) (blue solid line), and using TRM (black dotted line). The main differences in the estimated PDFs are due to the non-parametric nature of TRM. In particular, important differences can be observed between $15$ and $25 \;\text{km/s}$ and around  $30 \;\text{km/s}$ since the KDE utilized in TRM approximates $f_Y(y|\beta)$ as closely as possible to the histogram of the data. In addition, we note that when estimated PDF by TRM is used, has its maximum toward larger values of rotational speed. This could be due to a TRM error in detecting the most probable speed, as we have already observed in the numerical simulation (see Section 4) mainly for small number of samples.

% In addition, TRM could yield negative values if it not used carefully, which are not valid for a PDF.
\begin{figure*}[tb]
	\centering
	\subfigure{\includegraphics[width=0.3\textwidth]{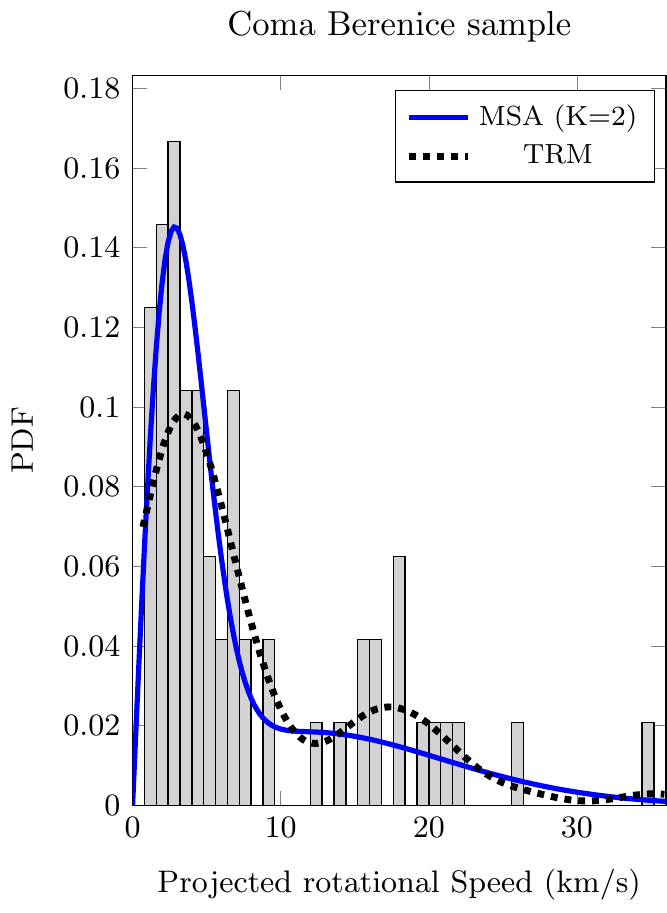}}
	\hfill
	\subfigure{\includegraphics[width=0.3\textwidth]{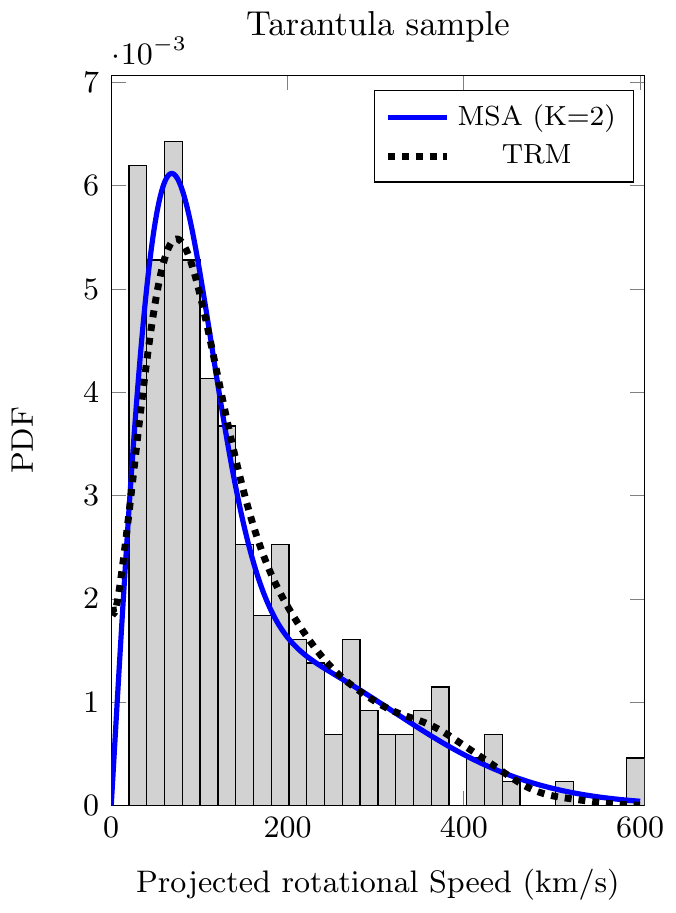}}
	\hfill
	\subfigure{\includegraphics[width=0.3\textwidth]{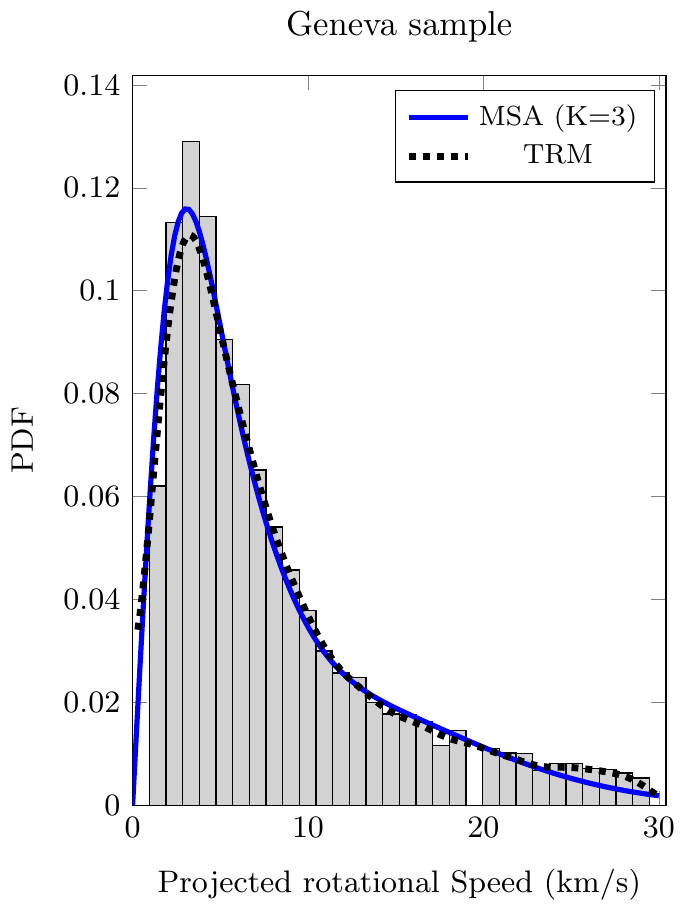}}
	%\centering
	%\includegraphics[width=1\textwidth,height=1\columnwidth,keepaspectratio,trim=29mm 0.78mm 29mm 0.5mm, clip]{Fig5_V10_2.eps}
	%\subfigure[]{\includegraphics[width=1\textwidth,height=1\columnwidth,keepaspectratio,trim=25mm 0.65mm 28mm 0.6mm, clip]{Fig5_a_V10.eps}}
	%\subfigure[]{\includegraphics[width=1\textwidth,height=1\columnwidth,keepaspectratio,trim=25mm 0.65mm 28mm 0.6mm, clip]{Fig5_b_V10.eps}}
	\caption{Contrasting projected and observed rotational velocities for the real samples using $K=2$ (Coma Berenice and Tarantula samples) and $K=3$ (Geneva sample) for MSA algorithm. Gray bars correspond to Histograms of the projected rotational velocities for the real samples. The blue line represents the corresponding estimated $f_{Y}(y|\beta)$ using MSA. The black dashed-dotted line represents the corresponding estimated $f_{Y}(y|\beta)$ using TRM.}
	\label{fig:Realsamples_PDFy_estimation}
\end{figure*}
\subsection{Tarantula sample}

We selected the Tarantula sample for single O-type stars form the VLT Flames Tarantula Survey presented in \cite{Agudelo2013}, where the authors deconvolved the rotational velocity distribution using the method in \cite{Lucy1974} and TRM. This sample contains 216 stars with $v\sin i$ data from $40\;\text{km/s}$ up to $610\;\text{km/s}$.

The estimated parameter using the MSA algorithm with one Maxwellian distribution ($K=1$) is $\sigma_1=131.39$. The parameters estimated for the two Maxwellian mixture distribution ($K=2$) are $\sigma_1=64.54$, $\sigma_2=185.67$, $\lambda_1=0.568$ and $\lambda_2=0.432$. The estimation results using three Maxwellian mixture distribution ($K=3$) are  $\sigma_1=64.54$, $\sigma_2=185.67$, $\sigma_3=64.54$, $\lambda_1=0.568$, $\lambda_2=0.432$ and $\lambda_3=8.03\times 10^{-17}$. It is evident that the mixing weight of the third component ($\lambda_3$) is not relevant in the Maxwellian mixture distribution model. In this sense, we focus on the results we obtained with $K=2$. Figure \ref{fig:Realsamples_PDFx_estimation}, middle panel, shows the rotational velocity distribution using MSA algorithm with $K=2$ (blue solid line) and the estimation using TRM (black dotted line). Similarly to the results from Coma Berenice sample, the estimated PDFs are similar, but exhibit some differences, particularly around $380$ to $480 \;\text{km/s}$, due to the use of KDE in TRM.

\subsection{Geneva sample}

We selected a large sample data of measured $v\sin i$ data of the Geneva-Copenhagen survey of the solar neighborhood (\cite{Nordstrom2004}, \cite{Holmberg2007}), which contains information about 16.500 F and G main-sequence field stars. We observed that this data sample presents important uncertainties showing velocities of $0\;\text{km/s}$, in consequence, we only selected stars with $0< v\sin i \leq 30\;\text{km/s}$, obtaining a sample of $11685$ stars.

The estimated parameter using MSA algorithm with one Maxwellian distribution is $\sigma_1=7.13$. Additionally, the estimated parameters for $K=2$ are $\sigma_1=3.32$, $\sigma_2=10.26$, $\lambda_1=0.578$ and $\lambda_2=0.422$. The estimation results using three Maxwellian mixture distribution are $\sigma_1=4.21$, $\sigma_2=10.66$, $\sigma_3=2.41$, $\lambda_1=0.40$, $\lambda_2=0.37$ and $\lambda_3=0.23$. In this case, we focus on the estimation obtained with three mixture components. Figure \ref{fig:Realsamples_PDFx_estimation}, right panel, shows the rotational velocity distribution using MSA algorithm with three Maxwellian mixture distribution (blue solid line) and TRM (black dotted line). In this case, we obtain an important agreement between the MSA estimation with $K=3$ and the TRM algorithm estimation, except for the interval between $25$ and $35\, \text{km/s}$.

\vspace{-2mm}
\subsection{Comparison of the estimation results}

In this section we describe our analysis of the estimation of the true rotational velocities for three data sets: Coma Berenice, Tarantula and Geneva, see Fig. \ref{fig:Realsamples_PDFx_estimation}. In addition, we considered the following: from the estimated true PDF of the rotational velocity, using MSA and TRM, we obtained an estimation of the projected rotational velocity PDF by solving the integral in Eq.\eqref{eq:integral_y_infnty}, in other words,
\begin{align}
\hat{f}_{Y}(y) = \int_{y}^{\infty}\frac{y}{x\sqrt{x^2-y^2}}\hat{f}_{X}(x)dx,
\end{align}
where $\hat{f}_X(x)=f_X(x|\hat{\beta})$ for MSA and $\hat{f}_X(x)$ is the estimated density using TRM, shown in Fig. \ref{fig:Realsamples_PDFy_estimation}.

From Fig. \ref{fig:Realsamples_PDFx_estimation} from a large number of samples (Tarantula and Geneva samples) both methods show similar results, except for at very low rotational speeds, where TRM wrongly estimates a non-zero probability for nought rotational velocities, that is, $\hat{f}_X(x=0)>0$. On the contrary, our method correctly provides a zero probability for nought rotational velocities, that is, $\hat{f}_X(x=0)=0$. In addition, the estimated projected rotational velocities are expected to provide a zero probability to nought projected rotational velocities. Our method provides the correct estimation (see Eq. \eqref{fy_Mtrue} and Eq. \eqref{fy_2Mtrue}), as is shown  in Fig. \ref{fig:Realsamples_PDFy_estimation}. On the other hand, similarly to the true rotational velocities estimation, the result provided by TRM is incorrect for zero projected rotational velocities.
For small number of samples (i.e., Coma Berenice sample), the true rotational velocity (Fig. \ref{fig:Realsamples_PDFx_estimation}) and the projected rotational velocity (Fig. \ref{fig:Realsamples_PDFy_estimation}) PDFs differ. However, the estimated projected rotational velocities we obtained with our method closely resembles the histogram from collected data (see Fig. \ref{fig:Realsamples_PDFy_estimation}). This results suggest that our method provides better estimate of the rotational velocity.

To validate our conclusion, we performed three statistical tests, a Kolmogorov-Smirnov (KS) test, an Anderson-Darling (AD) test and a q-q plot utilizing the measurements and the estimated projected rotational velocities. All three show that our method provides more accurate estimates of the true rotational velocities from the collected data for the three sets of samples. In particular, as is expected, for a large number of samples (i.e., Tarantula and Geneva samples), the results from the test are similar. For a small number of samples (i.e., Coma Berenice sample), our method has resulted in less discrepancy between the PDFs of the estimated projected rotational velocities and the one obtained form the measurements.

\begin{figure}
	\centering
	\includegraphics[width=0.48\textwidth]{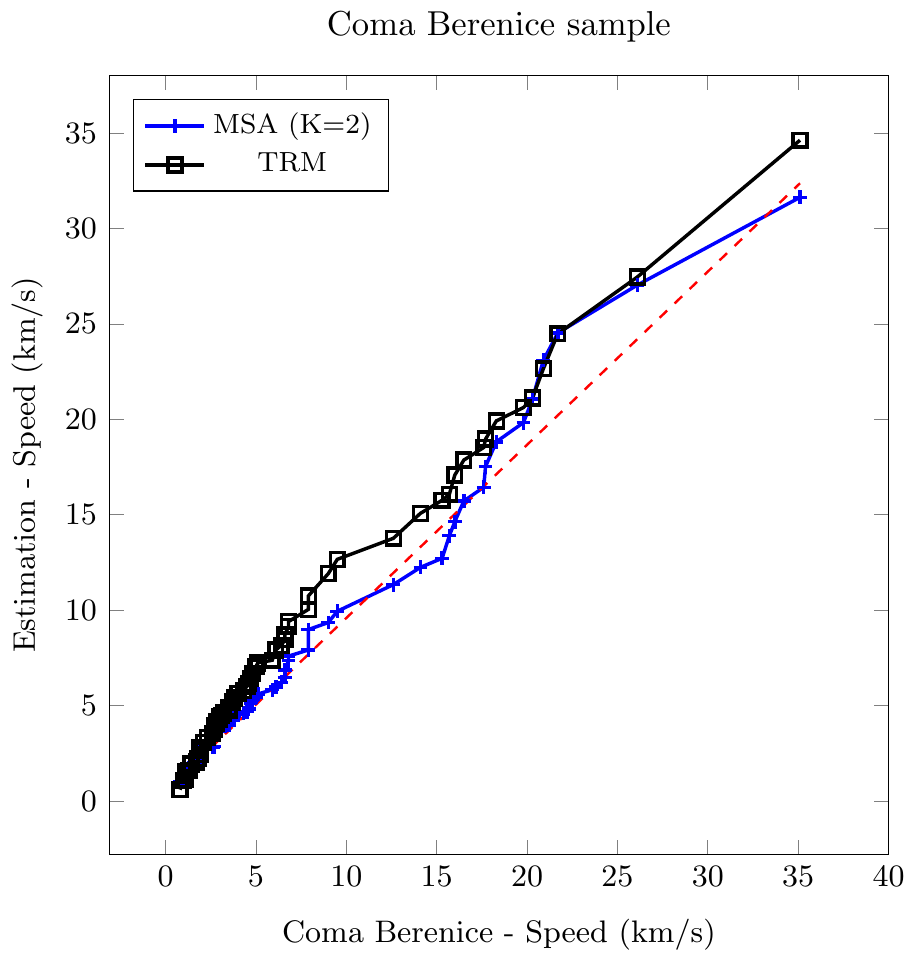}
	%\centering
	%\includegraphics[width=1\textwidth,height=1\columnwidth,keepaspectratio,trim=29mm 0.78mm 29mm 0.5mm, clip]{Fig5_V10_2.eps}
	%\subfigure[]{\includegraphics[width=1\textwidth,height=1\columnwidth,keepaspectratio,trim=25mm 0.65mm 28mm 0.6mm, clip]{Fig5_a_V10.eps}}
	%\subfigure[]{\includegraphics[width=1\textwidth,height=1\columnwidth,keepaspectratio,trim=25mm 0.65mm 28mm 0.6mm, clip]{Fig5_b_V10.eps}}
	\caption{q-q plot from Coma Berenice sample. The blue-mark line represents the quantiles of the corresponding estimated $f_{Y}(y|\beta)$ using MSA. The black-square line represents the quantiles of the corresponding estimated $f_{Y}(y|\beta)$ using TRM. The red-dashed line is the reference line.}
	\label{fig:qqplot_comaberenice}
\end{figure}

We used the KS test to support this analysis, where the not-rejection significance level is given as a percentage-the lower it is, the more reliable is the estimated model. We consider the empirical PDF from Coma Berenice sample and its corresponding CDF, obtained via numerical integration (trapezoidal method) using the estimated PDF from TRM and MSA algorithms. We tested performance between the real samples and the estimation results for the null hypothesis $H_0$ ``the projected velocities sample comes from the projected estimated distribution (from MSA or TRM)", and the complementary alternative hypothesis. Considering the MSA algorithm, we observed that the null hypothesis is not rejected at a significance level of $7.4\%$, and from TRM algorithm, the null hypothesis is not rejected at a significance level of $39.6\%$. Then, we also obtained the q-q plot (Fig. \ref{fig:qqplot_comaberenice}) of the densities estimated using TRM and MSA algorithms, confirming that MSA estimation closely fall on the reference line with respect to TRM results.

In order to corroborate our findings, we considered the AD test (see e.g., \cite{Anderson1952}),  considering a significance level of $5\%$. From the Coma Berenice sample, the null hypothesis is not rejected when the MSA algorithm is used, and the null hypothesis is rejected for TRM results. This might imply that the estimated PDF from TRM does not represent the actual data. From the Tarantula and Geneva samples, the null hypothesis is not rejected when we use the two algorithms (TRM and MSA) to estimate the corresponding $f_{Y}(y|\beta)$.

\vspace{-3mm}

\section{Conclusions and final remarks}

In this work we proposed a novel probability density function estimation algorithm using ML approach in terms of finite distribution mixtures, particularly, Maxwellian distributions. This algorithm is utilized to obtain the estimated probability distribution of true rotational stellar velocities. The advantage of this algorithm is that we were able to use the sample data from the experiments to obtain an estimated PDF of the projected rotational velocities $f_Y(y)$. This algorithm allowed us to use the sample data directly, without intermediate steps or approximations, unlike the estimation using TRM that requires the utilization of KDE, a non-parametric estimation technique.

We analyzed the performance of the proposed algorithm utilizing synthetic data and several Monte Carlo simulations, when the rotational velocity is described by: i) a Maxwell distribution and ii) a mixture of two or three Maxwellian distributions, considering different sample lengths ($N=50$, $N=500$ and $N=2000$). In general, we observed satisfactory results for each situation under different scenarios using MSA algorithm to estimate the true rotational velocity, exhibiting a better overall performance when compared to TRM. Moreover, we use the MISE as a measure of the performance of the PDF estimation. We obtained excellent results with MSA algorithm for all sample sizes, while the performance of TRM is satisfactory only for samples lengths of sizes $N=500$ and $N=2000$.

The PDF estimation algorithm was also tested in a set of real observed data from the Coma Berenice, Tarantula clusters and Geneva sample. We note that TRM has a lack of smoothing, likely due to the use of KDE. In addition, when utilizing TRM, the mode of the estimated function is slightly shifted to the right, compared to the estimated PDF from the proposed MSA. On the other hand, when using the proposed MSA algorithm with the Tarantula samples, the estimated PDF (for $K=2,3$) is very similar to the estimated function when using TRM method. This agreement in the estimates is due to the larger length of samples ($N \approx 500$). Finally, from the Geneva sample we obtained an important agreement between estimations from MSA with $K=3$ and TRM algorithms due to large length of samples used $N>2000$. %Finally, we obtain the 
% Moreover, the utilization of TRM with the samples from Coma Berenice cluster yielded a function with negative values; hence, the estimated function is not a PDF.
The proposed methodology can also be utilized for the estimation of other distributions associated with the projected rotational velocity. We have tested this idea with simulated data for the projected rotational velocity drawn from Tsallis distribution \cite{Carvalho2009}, Kanadiakis distribution \cite{Carvalho2009}, and Gamma-Normal distribution \cite{Agudelo2013}. The results yielded an important agreement in the statistical description of the projected rotational velocities, particularly with $K=4$ Maxwellian terms in the MSA. If more complex PDFs are to be estimated, other distributions can be used in the finite mixture distribution, for example Gaussian, provided they satisfy the Wiener approximation theorem \cite{ref:Achieser} in the desired domain of the PDF.

To summarize, we proposed a PDF estimation algorithm for rotational velocities based on ML, utilizing a finite mixture distribution that can be used even for small length of samples. The TRM proposed by \cite{Christen2016} exhibits a similar performance, but for a larger length of samples.
% since it requires the computation of a KDE. 

\begin{acknowledgements}
This work was partially supported by FONDECYT trough grant No 1181158, the Advanced Center for Electrical and Electronic Engineering (AC3E, Proyecto Basal FB0008), N\'ucleo Milenio de Formaci\'on Planetaria - NPF and Programa de Iniciaci\'on a la Investigaci\'on Cient\'ifica (PIIC) de la Direcci\'on de Postgrado UTFSM convenio No 015/2018. Michel Cur\'e thanks the support from the Centro de Astrof\'isica de Valpara\'iso and the Centro Interdisciplinario de Estudios Atmosf\'ericos y Astroestad\'istica.
\end{acknowledgements}
\vspace{-6mm}
   \bibliographystyle{aa} % style aa.bst
   \bibliography{biblio} % your references Yourfile.bib

\begin{appendix} %First appendix

	\section{Computing the parameters of the MSA}
	\label{appendix_B}
	We define the following
	\begin{align}
		&K(x_t,\beta_j)=\lambda_jg(x_t;\theta_j),\\
		&d\mu(x_t)=p(y_t|x_t)dx_t,
	\end{align}
	Then, the log-likelihood function in Eq. \eqref{eq:log_likelihood} can be expressed as
	\begin{equation}
		\ell_N(\beta)=\sum_{t=1}^{N}\log\left[\mathcal{V}_t(\beta)\right],
	\end{equation}
	with 
	\begin{equation}
		\label{eq_Vt}
		\mathcal{V}_t(\beta)=\sum_{j=1}^{K}\int_{y_t}^{\infty}K(x_t,\beta_j)d\mu(x_t),
	\end{equation}
	Thus, the ML estimator is obtained from
	\begin{equation}
		\label{eq_thetaML_hat}
		\hat{\beta}_\text{ML}=\arg \max_{\beta}\; \sum_{t=1}^{N}\log\left[\mathcal{V}_t(\beta)\right],
	\end{equation}
	By defining $\mathcal{B}_t(\beta)=\log\left[\mathcal{V}_t(\beta)\right]$, we can follow a similar analysis as in \cite{Carvajal2018}, obtaining
	\begin{equation}
		\mathcal{B}_t(\beta)=\mathcal{Q}_t(\beta,\hat{\beta}^{(m)})-\mathcal{H}_t(\beta,\hat{\beta}^{(m)}),
		\label{eq:Q_H}
	\end{equation}
	where
	\begin{align}
	\label{eq_Qauxiliar}
	&\mathcal{Q}_t(\beta,\hat{\beta}^{(m)})=\sum_{j=1}^{K}\int_{y_t}^{\infty}\log\left[K(x_t,\beta_j)\right]\frac{K(x_t,\hat{\beta}^{(m)}_j)}{\mathcal{V}_t(\hat{\beta}^{(m)})}d\mu(x_t),\\
	%\end{equation}
	%\begin{equation}
	&\mathcal{H}_t(\beta,\hat{\beta}^{(m)})=\sum_{j=1}^{K}\int_{y_t}^{\infty}  \! \log\left[\frac{K(x_t,\beta_j)}{\mathcal{V}_t(\beta)}\right]\frac{K(x_t,\hat{\beta}^{(m)}_j)}{\mathcal{V}_t(\hat{\beta}^{(m)})}d\mu(x_t).
	\end{align}
	Thus, from Eq. \eqref{eq_thetaML_hat} and Eq. \eqref{eq:Q_H}, the ML estimator can be locally obtained from
	\begin{equation}
	\label{eq_thetaML_hat_EM}
	\hat{\beta}_\text{ML}=\arg \max_{\beta}\; \sum_{t=1}^{N}\mathcal{B}_t(\beta),
	\end{equation}
	Then, the function $\mathcal{H}_t(\beta,\hat{\beta}^{(m)})$ is a decreasing function for any value of $\beta$ and satisfies the following:
	\begin{equation}
	\label{eq_H_decreasing}
	\mathcal{H}_t(\beta,\hat{\beta}^{(m)})-\mathcal{H}_t(\hat{\beta}^{(m)},\hat{\beta}^{(m)})\leq 0.
	\end{equation}
	In \cite{Orellana2018} is detailed the proof of this inequality. From this inequality inspired by the EM algorithm, we can formulate the following iterative algorithm:
	\begin{align}
	\label{eq_Estep}
	&\bar{\mathcal{Q}}(\beta,\hat{\beta}^{(m)})=\sum_{t=1}^{N}\mathcal{Q}_t(\beta,\hat{\beta}^{(m)}),\\
	\label{eq_Mstep}
	&\hat{\beta}^{(m+1)}=\arg \max_{\beta}\; \bar{\mathcal{Q}}(\beta,\hat{\beta}^{(m)}).
	\end{align}
	We note that Eq. (\ref{eq_Estep}) and Eq. (\ref{eq_Mstep}) correspond to the E-step and M-step of the EM algorithm, respectively.
	Taking derivative of $\bar{\mathcal{Q}}(\beta,\hat{\beta}^{(m)})$ with respect to $\alpha=1/\sigma^2_j$ and equating to zero we obtain
	\begin{equation}
	\begin{split}
	&\dparcial{\bar{\mathcal{Q}}(\beta,\hat{\beta}^{(m)})}{\alpha}=\frac{3}{2\hat{\alpha}^{(m+1)}}\sum_{t=1}^{N}\int_{y_t}^{\infty}\frac{K_M\left(x_t,\hat{\beta}_j^{(m)}\right)}{\mathcal{V}_{M_t}\left(\hat{\beta}^{(m)}\right)}d\mu(x_t)\\
	&-\frac12\sum_{t=1}^{N}\int_{y_t}^{\infty}x_t^2\frac{K_M\left(x_t,\hat{\beta}_j^{(m)}\right)}{\mathcal{V}_{M_t}\left(\hat{\beta}^{(m)}\right)}d\mu(x_t)=0
	\end{split}
	\end{equation}
	Using Eq. \eqref{Pxt_beta_MSA} and Eq. \eqref{Sxt_beta_MSA} we have
	\begin{equation}
	\begin{split}
	\left[{{\hat \sigma}^2_j}\right]{}^{\,\,(m+1)}&=\frac{1}{3\mathcal{P}_M(x_t,\hat{\beta}_j^{(m)})}\sum_{t=1}^{N}\int_{y_t}^{\infty}\!\!\!x_t^2\frac{K_M\left(x_t,\hat{\beta}_j^{(m)}\right)}{\mathcal{V}_{M_t}\left(\hat{\beta}^{(m)}\right)}d\mu(x_t)\\
	\left[{\hat{\sigma}_j}\right]{}^{\,\,(m+1)}&=\left[\frac{\mathcal{S}_M(x_t,\hat{\beta}_j^{(m)})}{3\mathcal{P}_M(x_t,\hat{\beta}_j^{(m)})}\right]^{\frac12}
	\end{split}
	\end{equation}
	For the parameter $\lambda_j$ we define $R_M(\lambda_j)$ as follows:
	\begin{equation}
	R_M(\lambda_j)=\sum_{j=1}^{K}\log\left[\lambda_j\right]\left\{\mathcal{P}_M(x_t,\hat{\beta}_j^{(m)})\right\}
	\end{equation}
	subject to
	\begin{equation}
	\sum_{j=1}^{K}\lambda_j=1
	\end{equation}
	Then, using Lagrange multipliers and optimizing with respect $\lambda_j$:
	\begin{equation}
	\mathcal{L}_M(\lambda_j,\gamma)=\sum_{j=1}^{K}\log\left[\lambda_j\right]\left\{\mathcal{P}_M(x_t,\hat{\beta}_j^{(m)})\right\}-\gamma_M\left(\sum_{j=1}^{K}\lambda_j-1\right)
	\end{equation}
	Taking the derivative of $\mathcal{L}_M(\lambda_j,\mu)$ with respect to $\lambda_j$ and $\gamma_M$, then equating to zero we obtain
	\begin{align}
	&\dparcial{\mathcal{L}_M(\lambda_j,\gamma_M)}{\lambda_j}=\frac{\mathcal{P}_M(x_t,\hat{\beta}_j^{(m)})}{\hat{\lambda}_j^{(m+1)}}-\gamma_M=0\\
	%\end{equation}
	%\begin{equation}
	\label{eq_sum_lamdaj_M}
	&\dparcial{\mathcal{L}_M(\lambda_j,\gamma_M)}{\gamma_M}=\sum_{j=1}^{K}\lambda_j-1=0.
	\end{align}
	Then,
	\begin{equation}
	\label{eq_lamdaj_lagrange_M}
	\hat{\lambda}_j^{(m+1)}=\frac{\mathcal{P}_M(x_t,\hat{\beta}_j^{(m)})}{\gamma_M}
	\end{equation}
	If we sum over $K$ in (\ref{eq_lamdaj_lagrange_M}) and use (\ref{eq_sum_lamdaj_M}) we have
	\begin{align}
	&\sum_{j=1}^{K}\hat{\lambda}_j^{(m+1)}=\sum_{j=1}^{K}\frac{\mathcal{P}_M(x_t,\hat{\beta}_j^{(m)})}{\gamma_M}=1\\
	%\end{equation}
	%\begin{equation}
	&\gamma_M=\sum_{j=1}^{K}\mathcal{P}_M(x_t,\hat{\beta}_j^{(m)})
	\end{align}
	Finally, we obtain
	\begin{equation}
	\begin{split}
	&\hat{\lambda}_j^{(m+1)}=\frac{\mathcal{P}_M(x_t,\hat{\beta}_j^{(m)})}{\sum_{l=1}^{K}\mathcal{P}_M(x_t,\hat{\beta}_l^{(m)})}
	\end{split}
	\end{equation}              
	%This completes the proof.
	\section{Estimation parameters from Monte Carlo simulations}
	The Table \ref{Tab:MSA_parameters_oneMaxwellian} shows the mean values and standard deviation of the estimation parameter of MSA algorithm using one Maxwellian mixture distribution. The case of study represents the unimodal distribution for true rotational velocities. %The Table \ref{Tab:GSA_parameters_oneMaxwellian} shows the results for the same case of study using GSA algorithm with two Gaussian mixture distribution.  
	\label{appendix_B3}
	\begin{table}[h]
		\centering
		\caption{Parameters estimated using MSA algorithm for one Maxwellian mixture distribution.}
		\label{Tab:MSA_parameters_oneMaxwellian}
		\begin{tabular}{cc}
			\hline \noalign{\smallskip}
			\begin{tabular}[c]{@{}c@{}}Parameters /\\ data length\end{tabular} & $\hat{\sigma}_1$ \\ \noalign{\smallskip} \hline \noalign{\smallskip}
			$50$  & $7.938 \pm 0.482$ \\ \noalign{\smallskip}
			$500$ & $8.001 \pm 0.177 $ \\ \noalign{\smallskip}
			$2000$ & $8.000 \pm 0.091$ \\ \hline \noalign{\smallskip}
		\end{tabular}
	\end{table}
	The Table \ref{Tab:MSA_parameters_twoMaxwellian} shows the mean values and standard deviation of the estimation parameter of MSA algorithm using two Maxwellian mixture distribution. The case of study represents the bimodal distribution for true rotational velocities. %The Table \ref{Tab:GSA_parameters_twoMaxwellian} shows the results for the same case of study using GSA algorithm with two Gaussian mixture distribution. 
	\begin{table*}[h]
		\centering
		\caption{Parameters estimated using MSA algorithm for two Maxwellian mixture distribution.}
		\label{Tab:MSA_parameters_twoMaxwellian}
		\begin{tabular}{ccccc}
			\hline \noalign{\smallskip}
			\begin{tabular}[c]{@{}c@{}}Parameters /\\ Data length\end{tabular} & $\hat{\lambda}_1$ & $\hat{\sigma}_1$ & $\hat{\lambda}_2$ & $\hat{\sigma}_2$ \\ \noalign{\smallskip} \hline \noalign{\smallskip}
			$50$  & $0.325 \pm 0.156$       & $5.05 \pm 1.85$       & $0.675 \pm 0.156$       & $15.15 \pm 1.98$       \\ \noalign{\smallskip}
			$500$ & $0.301 \pm 0.05$      & $5.01 \pm 0.48$       & $0.699 \pm 0.05$       & $15.05 \pm 0.48$       \\ \noalign{\smallskip}
			$2000$ & $0.298 \pm 0.02$      & $4.99 \pm 0.22$       & $0.702 \pm 0.02$       & $14.98 \pm 0.24$       \\ \hline \noalign{\smallskip}
		\end{tabular}
	\end{table*}
\section{Mean integrated square error (MISE) for Monte Carlo simulations}
	\label{appendix_B4}
The Table \ref{Tab:MISE_methods_unimodal} shows the MISE (Eq. \eqref{eq:MISE}) values for unimodal Maxwellian estimation considering one component ($K=1$) for MSA algorithm and the TRM algorithm for different sample data length. Similar, the Table \ref{Tab:MISE_methods_Bimodal} shows the MISE values for bimodal Maxwellian estimation considering one component ($K=2$) for MSA algorithm and the TRM algorithm different sample data length.
	\begin{table}[]
		\centering
		\caption{MISE values of PDF estimation for unimodal Maxwellian distribution example.}
		\label{Tab:MISE_methods_unimodal}
		\begin{tabular}{ccc}
			\hline    \noalign{\smallskip}
			\multirow{2}{*}{\begin{tabular}[c]{@{}c@{}}Data length /\\ Method \end{tabular}} & \multicolumn{2}{c}{MISE} \\ \cline{2-3}    \noalign{\smallskip}
			& $MSA$           & $TRM$   \\ \hline        \noalign{\smallskip}
			$50$                                                                        &   $5.07 \times 10^{-6}$            &  $9.14 \times 10^{-5}$         \\              \noalign{\smallskip}
			$500$                                                                        &    $4.72 \times 10^{-6}$         &   $1.68 \times 10^{-5}$       \\                \noalign{\smallskip}
			$2000$                                                                              &       $4.71 \times 10^{-6}$       &     $6.59 \times 10^{-6}$   \\ \hline
		\end{tabular}
	\end{table}

	\begin{table}[t]
		\centering
		\caption{MISE values of PDF estimation for bimodal Maxwellian distribution example.}
		\label{Tab:MISE_methods_Bimodal}
		\begin{tabular}{ccc}
			\hline    \noalign{\smallskip}
			\multirow{2}{*}{\begin{tabular}[c]{@{}c@{}}Data length /\\ Method \end{tabular}} & \multicolumn{2}{c}{MISE} \\ \cline{2-3}    \noalign{\smallskip}
			& $MSA$            & $TRM$ \\ \hline        \noalign{\smallskip}
			$50$                                                                        &   $7.59 \times 10^{-6}$          &    $5.05 \times 10^{-5}$       \\              \noalign{\smallskip}
			$500$                                                                        &    $7.52 \times 10^{-6}$         &    $1.06 \times 10^{-5}$ \\                \noalign{\smallskip}
			$2000$                                                                              &       $7.48 \times 10^{-6}$       &    $7.49 \times 10^{-6}$ \\ \hline
		\end{tabular}
	\end{table}

\end{appendix}
\end{document}